\documentclass[preprint,aps,prl,preprintnumbers,amsmath,amssymb,superscriptaddress,showpacs,floatfix]{revtex4}
\usepackage{epsfig}
\usepackage{dcolumn}
\usepackage{bm}
\usepackage{amssymb}
\usepackage[bookmarks=false]{hyperref}
\newcommand{\order}[1]{ \mathcal{O} \left( #1 \right) }

\begin{document}

\title{Transport Properties of QCD at Large $N_c$ and the Gauge/String Duality}

\author{Jorge Noronha}
\affiliation{Department of
Physics, Columbia University, 538 West 120$^{th}$ Street, New York,
NY 10027, USA}

\begin{abstract}

Below the deconfinement phase transition large $N_c$ QCD is expected to be a very viscous hadronic fluid because both the shear and bulk viscosity to entropy density ratio, $\eta/s,\zeta/s \sim N_c^2$. In this letter I show that $\eta/s \sim N_c^2$ in the confined phase of holographic models of QCD at large $N_c$ defined in the supergravity approximation. Our results show that the gauge/string duality can be used to describe not only nearly perfect fluids but also extremely viscous systems such as a cold gas of glueballs. 

\end{abstract}

\maketitle


Lattice calculations \cite{largeNclattice} have shown that the thermodynamic properties of $SU(N_c)$ Yang-Mills (YM) theories do not change significantly above $T_c$ if $N_c \geq 3$. The deconfinement transition remains first-order and all the thermodynamic quantities, when properly rescaled by $N_c^2$, seem to display a universal behavior when $N_c\geq 3$. This supports the idea that the large $N_c$ approximation is a good starting point in the study of the equilibrium properties of YM theories and perhaps also the RHIC plasma. It is also natural to assume that the large $N_c$ approximation can be useful to compute the transport properties of non-Abelian gauge theories.  

Indeed, strongly-coupled gauge theories at large $N_c$ constructed within the Anti-de Sitter/Conformal Field Theory (AdS/CFT) correspondence \cite{Maldacena:1997re,AdSCFTcorrelators,Aharony:1999ti} seem to already capture some important aspects of the RHIC data such as the large heavy quark energy loss \cite{heavyquarkenergyloss} and the nearly perfect fluidity evidenced by the strong elliptic flow \cite{RHIC}, which suggests that the deconfined medium is some type of strongly-coupled liquid (the sQGP) \cite{sQGP} where, on average, $4\pi\,\eta/s \sim 1$ \cite{Luzum:2008cw}. Note that after hadronization the matter is expected to be considerably more viscous at low temperatures \cite{viscoushadrongas,csernai} although near $T_c$ effects from Hagedorn resonances may be able to lower $\eta/s$ to a value similar to the one in the sQGP \cite{jaki}. 

It has been shown that $\eta/s=1/(4\pi)$ in the deconfined phase of any gauge theory dual to supergravity \cite{Kovtun:2004de}. Even though the deconfined degrees of freedom in these strongly-coupled gauge theories do not admit a quasiparticle interpretation, the general idea from kinetic theory that strong coupling leads to small mean free path is somehow valid in this case. If higher-order derivative corrections are included in the gravity action $\eta/s$ changes but it remains $\sim 1$ \cite{myers} (from now on $\sim 1$ means that the quantity does not depend on $N_c$ to leading order). At high temperatures, perturbative QCD calculations predict that $\eta/s \sim 1$ \cite{Arnold:2000dr}. 

However, below the deconfinement phase transition colorless states are the relevant degrees of freedom and one expects that $\eta/s \sim N_c^2$ \cite{csernai}. In fact, in QCD at very low temperatures momentum isotropization is done mostly by pions and in the chiral limit $\eta/s \sim f_{\pi}^4/T^4 \sim N_c^2$ \cite{prakash}, where $f_{\pi} \sim \sqrt{N_c}$ is the pion decay constant. Furthermore, the bulk viscosity of a gas of massive pions shows a similar behavior $\zeta/s \sim N_c^2$ \cite{FernandezFraile:2008vu}. 

In the confined phase of YM at large $N_c$, glueballs are expected to be well defined (non-relativistic) quasiparticles with negligible width that interact very weakly \cite{wittenlargeNc}. Under such conditions, standard kinetic theory \cite{Danielewicz:1984ww} predicts that $\eta\sim  \langle p \rangle \, n\, \ell_{MFP}$, where the typical momentum is $\langle p\rangle \sim \sqrt{M\,T}$ (here $M\gg T$ is the glueball mass), $n$ is the density, and $\ell_{MFP} \sim N_c^2$ is the mean free path. Indeed, at large $N_c$ a gas of glueballs is so dilute and weakly interacting that its properties at (and near) equilibrium should be perfectly described by kinetic theory.  

While it is by now clear that the deconfined phase of strongly-coupled gauge theories with gravity duals exhibits nearly perfect fluid behavior, it is not at all evident that strongly interacting models based on holography can also describe the extremely viscous and dilute properties of the confined phase of gauge theories at large $N_c$. In this letter I argue that the effective colorless degrees of freedom in the confined phase of holographic models of QCD at large $N_c$ become extremely weakly interacting at low $T$ by showing that $\eta/s \sim N_c^2$. This indicates that holographic methods can also be used to understand the near-equilibrium properties of systems that are usually described within kinetic theory.  

In YM at large $N_c$ one can expand $\eta$ and $s$ as $\eta=\eta_2\,N_c^2+\eta_0 +\order{1/N_c^2}$, $s=s_2\,N_c^2 + s_0 +\order{1/N_c^2}$, where the coefficients $\eta_i$, $s_i$ depend on $T$ and also the t'Hooft coupling $\lambda=g^2_{YM}N_c$. In the confined phase $s_2(T<T_c)=0$ while $\eta_2(T<T_c)$ is nonzero and, thus, $\eta/s=N_c^2\,\eta_2/s_0$ plus subleading corrections that vanish at large $N_c$. When $T/M_0 \ll 1$ (with $M_0$ being the mass of the lightest glueball) one obtains that
\begin{equation}
\ s_0(T) \sim \frac{M_0}{T} \left(\frac{M_0\,T}{2\pi}\right)^{3/2}\,e^{-M_0/T}\,.
\label{sglueball}
\end{equation}    
This approximation gives the smallest $s_0$ and hence the largest value of $\eta/s$ in the confined phase. 

At strong coupling, the deconfinement phase transition is understood as a Hawking-Page transition \cite{Hawking:1982dh} from a geometry without a horizon (thermal AdS $\to$ confined phase) to another one where a horizon exists (black brane $\to$ deconfined phase) \cite{Witten:1998zw}. The transition is then first-order with a specific heat of order $N_c^2$. Here we assume that the background geometry corresponding to the confined phase is gapped and $s_0$ at low $T$ is given by Eq.\ (\ref{sglueball}). Note that while $s_0$ in the confined phase is obtained via a one-loop correction in the gravity dual, the leading contribution to $\eta\sim \eta_2\,N_c^2$ at large $\lambda$ is fully determined by classical supergravity. 

Some well-known holographic models for the confined phase include bottom-up models such as the hard \cite{hardwall} and soft wall \cite{softwall} models and also top-down constructions such as the Sakai-Sugimoto model \cite{Sakai:2004cn}. The deconfinement phase transition in these models is a Hawking-Page transition \cite{Aharony:2006da,Herzog:2006ra} and the confined phase displays a mass gap. However, in the hard wall and Sakai-Sugimoto models the masses of very excited hadrons do not follow the observed linear Regge behavior $M_n^2 \sim n$, where $n \gg 1$ is the radial excitation number. While in the soft wall model \cite{softwall} linear confinement is built in, the background metric in the string frame is identical to $AdS_{5}$ and, thus, the Wilson loop does not obey an area law.

Here we will consider holographic models of the type proposed by Gursoy and collaborators \cite{gursoy} who showed that several properties of YM theories at large $N_c$ can be obtained via the 5d gravity action (in the Einstein frame)
\begin{equation}
\mathcal{A}=\frac{1}{16\pi G_5} \int d^5x\sqrt{-G}\left[\mathcal{R}-\frac{4}{3}(\partial \Phi)^2-V(\Phi)\right]\,,
\label{dilatonaction}
\end{equation}         
which describes the interactions between gravity $G_{\mu\nu}$ and a dilaton-like field $\Phi$. The t'Hooft coupling in the gauge theory is assumed here to be $\lambda=g_{YM}^2\,N_c \equiv e^{\Phi}$. The basic assumption behind such models is that YM at large $N_c$ should be described by a noncritical 5d string theory where the bulk fields $G_{\mu\nu}$ and $\Phi$ are dual to $T_{\mu\nu}$ and ${\rm Tr} F^2$ in the gauge theory. The axion in the bulk is dual to ${\rm Tr} F \wedge F$ but its contribution to the action is subleading at large $N_c$. The effects of the other fields in the bulk are assumed to be somehow included in the dilaton potential $V(\Phi)$. The potential used in \cite{gursoy} is such that the geometry approaches $AdS_5$ logarithmically in order to mimic effects from asymptotic freedom.  

It was shown in \cite{gursoy} that a first-order phase transition that closely resembles the deconfining transition in YM at large $N_c$ can be obtained if in the IR (where $\Phi \gg 1$) the potential becomes 
\begin{equation}
\ V(\Phi) \sim - e^{4\Phi/3} \,\sqrt{\Phi}\,.
\label{potential}
\end{equation}
Below the critical temperature $T_c$ the background fields are the vacuum solutions 
\begin{equation}
\ ds^2 = e^{2A_0(r)}\left(-dt^2+d\vec{x}^{\,2}+dr^2\right), \qquad \Phi=\Phi_0(r)\,.
\label{confined}
\end{equation}
Finite temperature effects are included in Euclidean space $t\to i\,\tau$, where $\tau$ is a periodic function $\tau \to \tau + 1/T$. This horizonless geometry corresponds to the thermal $AdS$ solution discussed by Hawking and Page in \cite{Hawking:1982dh} where the boundary is located at $r=0$. It is easy to see that the leading contribution to the entropy density ($\sim N_c^2$) vanishes in this geometry since the system's free energy $\mathcal{F}=T\, \mathcal{A}$ does not depend on $T$. A potential with IR behavior given by (\ref{potential}) leads to a linearly confined theory with glueball masses that are consistent with current lattice data \cite{gursoy}. 

In hard wall-like models confinement is associated with a scale in the bulk, say, $r_\sigma$, where spacetime literally ``ends" \cite{hardwall}. This scale also determines the maximum depth reached by the dual string between heavy quark probes at the boundary that experience a linear (hence confining) potential. A similar type of singularity below $T_c$ is also present in the model (\ref{dilatonaction}) when $r\to \infty$ (this is in fact a general property of these two-derivative actions \cite{gursoy}) but this singularity can only be probed by highly excited states while classical string worldsheets remain far from this region where the curvature diverges. Above $T_c$, the singularity is shielded by the black brane horizon. The maximum depth, $r_\sigma$, probed by a classical string is determined dynamically using the classical solutions of the string equations of motion in the background (\ref{confined}) and the confining string tension is     
\begin{equation}
\sigma =T_s e^{2A_0(r_\sigma)}\lambda_\sigma^{4/3}
\label{stringtension}
\end{equation}  
where $\lambda_{\sigma}=e^{\Phi_0(r_\sigma)}$ and $T_s=1/(2\pi\ell_s^2)$ is the string tension defined in terms of the fundamental string length. 

The deconfined phase involves a black brane 
\begin{equation}
\ ds^2 = e^{2A(r)}\left(-f(r)dt^2+d\vec{x}^{\,2}+\frac{dr^2}{f(r)}\right), \qquad \Phi=\Phi(r)\,.
\label{deconfined}
\end{equation}
with thermodynamic properties \cite{gursoy} that match lattice calculations \cite{Boyd:1996bx}. The horizon function $f(r)$ has a simple root at the horizon's location $r_h$. Similar black brane solutions where investigated in \cite{nellore} but in these studies the dilaton-like field was assumed to be dual to an operator of dimension $\Delta <4$ in the boundary. These gravity duals are different than those considered in \cite{gursoy} because they have a nontrivial UV fixed point where the geometry becomes $AdS_5$ with radius $R$ and the potential assumes the universal form $V(\Phi) \sim -12 +m_{\Phi}^2 \Phi^2/2$, where $m_{\Phi}=\Delta(\Delta-4) < 0$. In this case, $\Phi$ is dual to ${\rm Tr}F^2$ at a finite scale in the IR above which the system is conformal invariant and not asymptotically free. However, the UV behavior is not relevant in our discussion and, in fact, it is sufficient for our purposes here to assume that the confined phase is of the form (\ref{confined}) with a string tension given by (\ref{stringtension}).

The shear viscosity in the field theory is computed using Kubo's formula
\begin{eqnarray}
\eta &=& \lim_{\omega\to 0}\lim_{\vec{k}\to 0}\int d^4x\,\frac{e^{-i\,\omega t}}{2\omega}\langle [T_{xy}(t,\vec{x}),T_{xy}(0)] \rangle \\
&=& -\lim_{\omega\to 0}\lim_{\vec{k}\to 0}\frac{1}{\omega}{\rm Im}\,G^{R}(\omega,\vec{k})
\label{kubo}
\end{eqnarray}
where $G^{R}$ is the retarded correlator of $T_{xy}$ in the field theory. The graviton absorption cross section for extremal black branes was computed in \cite{klebanov}. According to the AdS/CFT correspondence, the energy-momentum tensor is the conserved current that couples to the metric disturbances $G_{\mu\nu}\to G_{\mu\nu}+\delta h_{\mu\nu}$ at the boundary of the bulk spacetime. In general, the line element is assumed to have the diagonal form
\begin{equation}
\ ds^2 = -G_{00}(r)dt^2+G_{xx}(r)d\vec{x}^2+G_{rr}(r)dr^2 \,.
\label{metric1}
\end{equation}
The fluctuation of the metric relevant to $\eta$ is given by $\delta h_{xy}$ and in the supergravity approximation the quadratic part of the graviton action becomes 
\begin{equation}
\ -\frac{1}{32 \pi G_5}\int d^5x\,\sqrt{-\det G}\,G^{\mu\nu}\,\partial_{\mu}\chi\,\partial_{\nu}\chi
\label{einstein1}
\end{equation}    
where $\chi\equiv \delta h_{y}^x$. In theories with higher-order corrections a similar action for $\chi$ can be found but the gravitational coupling is in general different than $G_5$ \cite{Brustein:2008cg}. The shear viscosity can be obtained from the canonical momentum $\Pi(x,r)$ associated with the solution of the equations of motion for $\chi(x,r)$ \cite{Iqbal:2008by}
\begin{equation}
\Pi(x,r) = -\frac{\sqrt{-G}}{16\pi\, G_5}\,G^{rr}\partial_{r}\chi(x,r) 
\label{momentum}
\end{equation} 
as follows
\begin{equation}
\eta = \lim_{\omega\to 0}\lim_{r\to 0}{\rm Im}\,\left(\frac{\Pi(\omega,0,r)}{\omega\,\chi(\omega,0,r)}\right)\,.
\label{eta1}
\end{equation} 

We shall assume in the following that the disturbance does not depend on the spatial momentum and also $\chi(\omega,r)=\chi_0(\omega)g_{\omega}(r)$ where $g_{\omega}(r)$ obeys the following equation
\begin{equation}
\left[\left(\sqrt{G_{00}\,G^{rr}\,G_{xx}^3}\partial_r\right)^2+\omega^2 G_{xx}^3\right]g_{\omega}(r)=0\,,
\label{EOM1}
\end{equation}
which is obtained from the equations of motion for $\chi$. It is convenient \cite{theoremcrosssection} to define the new coordinate $\xi(r)=\int^{r}d\tilde{r}/\sqrt{G_{00}\,G^{rr}\,G_{xx}^3}$ so then Eq.\ (\ref{EOM1}) becomes  
\begin{equation}
\left[\partial_\xi^2+\omega^2 G_{xx}^3(r(\xi))\right]\,g_{\omega}(\xi)=0\,.
\label{EOM2}
\end{equation}
An important property of Eq.\ (\ref{eta1}) is that in the limit $\omega\to 0$ (with $\omega \chi$ and $\Pi$ fixed) the same result for $\eta$ is found regardless of the radius used in the evaluation of the right-hand side in (\ref{eta1}), as shown in \cite{Iqbal:2008by}. This follows directly from the definition of $\Pi$ and its corresponding equations of motion. Thus, assuming that $\xi^*=\xi(r^*)$ is well defined at a given $r^* \in (0,\infty)$ and $G_{xx}(r^*)\neq 0$, near $r^*$ the general solution of (\ref{EOM2}) is a sum of incoming and outgoing waves at $r^*$ 
\begin{equation}
\ g_{\omega}(r)=A \, e^{-i\omega G_{xx}^{3/2}(r^*)\xi(r)}+ B \, e^{i\omega G_{xx}^{3/2}(r^*)\xi(r)}\,,
\label{solution1}
\end{equation}
which can be seen as a consequence of time-reversal invariance. However, in general at finite temperature one should expect that time-reversal invariance is broken and the retarded Green's function should be associated with the incoming wave at $r^*$, which leads to 
\begin{equation}
\eta = \frac{G_{xx}^{3/2}(r^*)}{16\pi\,G_5}\,.
\label{eta2}
\end{equation} 
In the deconfined phase the black brane contribution to the entropy is simply $G_{xx}^{3/2}(r_h)/(4G_5)$ and, after taking $r^*=r_h$, one immediately recovers the general result that $\eta/s=1/(4\pi)$ in the deconfined phase of all gauge theories dual to supergravity \cite{Kovtun:2004de}. 

We shall now compute $\eta$ in the background (\ref{confined}) that is assumed to describe the confined phase of YM at large $N_c$. Such geometries do not have a horizon but the general discussion presented above still applies. In fact, $\eta$ can also be computed at any radius $r^*$ where $\xi^*$ is well defined and $e^{A_0(r^*)}\neq 0$. Naively, one may think that in this case a knowledge of the full solution of the equations of motion for $A_0(r)$ is necessary to evaluate $\eta$. However, this is not true. In the confining geometries obtained as solutions of (\ref{dilatonaction}) $e^{A_0(r)}$ only vanishes at the singularity $r\to \infty$ where $A_0(r) \sim -r^2$ \cite{gursoy}. Moreover, evaluating $\eta$ at $r_\sigma$ and using (\ref{stringtension}) one obtains    
\begin{equation}
\eta(T<T_c) = \frac{1}{16\pi\,G_5}\,\left(\frac{\sigma}{T_s}\right)^{3/2}\frac{1}{\lambda_{\sigma}^2}\,.
\label{eta4}
\end{equation}
This is the main result of this letter. It shows that $\eta \sim N_c^2$ in holographic models of YM at large $N_c$, which indicates that these models may indeed provide the correct degrees of freedom below $T_c$. One should also expect that $\zeta \sim N_c^2$ in such geometries.

Other transport coefficients associated with shear dissipation can be found in the hydrodynamic expansion and to 2nd order in the gradients one finds $\tau_{\pi}$, which basically gives the amount of time necessary for a flow gradient to be converted into heat. Causality requires that $(\eta/s)/(\tau_{\pi}\,T)$ is finite and smaller than unity \cite{causalityreview}. Therefore, in order to preserve causality in the deconfined phase $\tau_{\pi} \,T$ must be at least of order one while in the confined phase it should scale with $\sim N_c^2$ or even a higher power. Indeed, for a very dilute gas with pressure $p$ one can show that $\tau_{\pi}=\eta/p$ \cite{Koide:2009sy}.

It is generally believed that $\eta/s$ has a minimum near $T_c$ \cite{csernai} and, because of the causality constraint, one should expect that $\tau_{\pi}T$ should also display a minimum near $T_c$. The fact that in the confined phase at large $N_c$ both $\eta/s \sim N_c^2$ and $\tau_{\pi} T$ $\sim N_c^2$ or higher indicates that the system is so dilute that, in the context of heavy ion collisions, kinetic freeze-out would basically occur exactly when the system hadronizes. In this limit, sQGP observables such as the elliptic flow coefficient become identically zero below $T_c$.   

One can use Eqs.\ (\ref{sglueball}) and (\ref{eta4}) to derive the upper bound $\eta/s \leq \eta(T<T_c)/s_0(T)$ for the confined phase of gravity duals. Thus, since this ratio becomes very large at large $N_c$ we see that the gauge/string duality can be used to describe not only nearly perfect fluids but also extremely viscous systems such as a gas of glueballs which, as argued above, should be perfectly described using a quasi-particle description derived within kinetic theory. The results discussed here should also be valid in the case of a more complex holographic theory where the confined phase involves not only glueballs but also mesons and hadrons. In this case, holographic methods could provide a reasonable framework to investigate the properties of the hot hadronic matter produced in relativistic heavy ion collisions.

I thank A.~Dumitru, M.~Gyulassy, G.~Torrieri, G.~Denicol, F.~Rocha, A.~Yarom, M.~Paulos, and J.~Casalderrey-Solana for discussions. This work was supported by the US-DOE Nuclear Science Grant No.\ DE-FG02-93ER40764 and the Helmholtz International Center for FAIR (LOEWE program).


\begin{thebibliography}{99}


\bibitem{largeNclattice}
B.~Bringoltz and M.~Teper,
  Phys.\ Lett.\  B {\bf 628}, 113 (2005); B.~Lucini, M.~Teper and U.~Wenger,
  JHEP {\bf 0502}, 033 (2005); M.~Panero,
  Phys.\ Rev.\ Lett.\  {\bf 103}, 232001 (2009).

\bibitem{Maldacena:1997re}
  J.~M.~Maldacena,
  Adv.\ Theor.\ Math.\ Phys.\  {\bf 2}, 231 (1998)
  [Int.\ J.\ Theor.\ Phys.\  {\bf 38}, 1113 (1999)].

\bibitem{AdSCFTcorrelators} 
S.~S.~Gubser, I.~R.~Klebanov and A.~M.~Polyakov,
  Phys.\ Lett.\  B {\bf 428}, 105 (1998); E.~Witten,
  Adv.\ Theor.\ Math.\ Phys.\  {\bf 2}, 253 (1998).

\bibitem{Aharony:1999ti}
  For a review see, O.~Aharony, S.~S.~Gubser, J.~M.~Maldacena, H.~Ooguri and Y.~Oz,
  Phys.\ Rept.\  {\bf 323}, 183 (2000).

\bibitem{heavyquarkenergyloss}
  S.~S.~Gubser,
  Phys.\ Rev.\  D {\bf 74}, 126005 (2006);  C.~P.~Herzog, A.~Karch, P.~Kovtun, C.~Kozcaz and L.~G.~Yaffe,
  JHEP {\bf 0607}, 013 (2006); J.~Casalderrey-Solana and D.~Teaney,
  Phys.\ Rev.\  D {\bf 74}, 085012 (2006); J.~Noronha, M.~Gyulassy and G.~Torrieri,
  arXiv:0906.4099 [hep-ph].

\bibitem{RHIC}
  I. Arsene et al. (BRAHMS), Nucl. Phys. A757, 1 (2005); B. B. Back et al., Nucl. Phys. A757, 28 (2005); J. Adams et al. (STAR), Nucl. Phys. A757, 102 (2005); K. Adcox et al. (PHENIX), Nucl. Phys. A757, 184 (2005).


\bibitem{sQGP} M.~Gyulassy and L.~McLerran,
  Nucl.\ Phys.\  A {\bf 750}, 30 (2005); E.~V.~Shuryak,
  Nucl.\ Phys.\  A {\bf 750}, 64 (2005).

\bibitem{Luzum:2008cw}
  H.~J.~Drescher, A.~Dumitru, C.~Gombeaud and J.~Y.~Ollitrault,
  Phys.\ Rev.\  C {\bf 76}, 024905 (2007); M.~Luzum and P.~Romatschke,
  Phys.\ Rev.\  C {\bf 78}, 034915 (2008)
  [Erratum-ibid.\  C {\bf 79}, 039903 (2009)].

\bibitem{viscoushadrongas}
T.~Hirano and M.~Gyulassy,
  Nucl.\ Phys.\  A {\bf 769}, 71 (2006); N.~Demir and S.~A.~Bass,
  Phys.\ Rev.\ Lett.\  {\bf 102}, 172302 (2009).

\bibitem{csernai}
L.~P.~Csernai, J.~I.~Kapusta and L.~D.~McLerran, Phys.\ Rev.\ Lett.\ {\bf 97}, 152303 (2006). 

\bibitem{jaki} J.~Noronha-Hostler, J.~Noronha and C.~Greiner,
  Phys.\ Rev.\ Lett.\  {\bf 103}, 172302 (2009).

\bibitem{Kovtun:2004de}
  P.~K.~Kovtun, D.~T.~Son and A.~O.~Starinets,
  Phys.\ Rev.\ Lett.\  {\bf 94}, 111601 (2005); A.~Buchel and J.~T.~Liu,
  Phys.\ Rev.\ Lett.\  {\bf 93}, 090602 (2004).

\bibitem{myers}
M.~Brigante, H.~Liu, R.~C.~Myers, S.~Shenker and S.~Yaida, Phys.\ Rev.\ D {\bf 77}, 126006 (2008); Phys.\ Rev.\ Lett.\
{\bf 100}, 191601 (2008); Y.~Kats and P.~Petrov, JHEP {\bf 0901}, 044 (2009); A.~Buchel, R.~C.~Myers
and A.~Sinha, JHEP {\bf 0903}, 084 (2009); R.~C.~Myers, M.~F.~Paulos and A.~Sinha, Phys.\ Rev.\
D {\bf 79}, 041901 (2009).

\bibitem{Arnold:2000dr}
  P.~Arnold, G.~D.~Moore and L.~G.~Yaffe,
  JHEP {\bf 0011}, 001 (2000); JHEP {\bf 0305}, 051 (2003).

\bibitem{prakash} M.~Prakash, M.~Prakash, R.~Venugopalan and G.~Welke, Phys.\ Rep.\ {\bf 227}, 321 (1993).

\bibitem{FernandezFraile:2008vu}
  D.~Fernandez-Fraile and A.~G.~Nicola,
  Phys.\ Rev.\ Lett.\  {\bf 102}, 121601 (2009).

\bibitem{wittenlargeNc}
  E.~Witten,
  Nucl.\ Phys.\  B {\bf 160}, 57 (1979).

\bibitem{Danielewicz:1984ww}
  P.~Danielewicz and M.~Gyulassy,
  Phys.\ Rev.\  D {\bf 31}, 53 (1985).

\bibitem{Hawking:1982dh}
  S.~W.~Hawking and D.~N.~Page,
  Commun.\ Math.\ Phys.\  {\bf 87}, 577 (1983).

\bibitem{Witten:1998zw}
  E.~Witten,
  Adv.\ Theor.\ Math.\ Phys.\  {\bf 2}, 505 (1998).

\bibitem{hardwall}   J.~Polchinski and M.~J.~Strassler,
  Phys.\ Rev.\ Lett.\  {\bf 88}.

\bibitem{softwall} A.~Karch, E.~Katz, D.~T.~Son and M.~A.~Stephanov,
  Phys.\ Rev.\  D {\bf 74}, 015005 (2006).

\bibitem{Sakai:2004cn}
  T.~Sakai and S.~Sugimoto,
  Prog.\ Theor.\ Phys.\  {\bf 113}, 843 (2005). 

\bibitem{Herzog:2006ra}
  C.~P.~Herzog,
  Phys.\ Rev.\ Lett.\  {\bf 98}, 091601 (2007).

\bibitem{Aharony:2006da}
  O.~Aharony, J.~Sonnenschein and S.~Yankielowicz,
  Annals Phys.\  {\bf 322}, 1420 (2007).

\bibitem{gursoy}
  U.~Gursoy and E.~Kiritsis,
  JHEP {\bf 0802}, 032 (2008); U.~Gursoy, E.~Kiritsis and F.~Nitti,
  JHEP {\bf 0802}, 019 (2008); U.~Gursoy, E.~Kiritsis, L.~Mazzanti and F.~Nitti,
  Phys.\ Rev.\ Lett.\  {\bf 101}, 181601 (2008); JHEP {\bf 0905}, 033 (2009); Nucl.\ Phys.\  B {\bf 820}, 148 (2009).


\bibitem{Boyd:1996bx}
  G.~Boyd, J.~Engels, F.~Karsch, E.~Laermann, C.~Legeland, M.~Lutgemeier and B.~Petersson,
  Nucl.\ Phys.\  B {\bf 469}, 419 (1996).

\bibitem{nellore} S.~S.~Gubser and A.~Nellore,
  Phys.\ Rev.\  D {\bf 78}, 086007 (2008); S.~S.~Gubser, A.~Nellore, S.~S.~Pufu and F.~D.~Rocha,
  Phys.\ Rev.\ Lett.\  {\bf 101}, 131601 (2008); A.~Cherman and A.~Nellore,
  Phys.\ Rev.\  D {\bf 80}, 066006 (2009); J.~Noronha,
  arXiv:0910.1261 [hep-th]. 
  

\bibitem{klebanov} 
  I.~R.~Klebanov, Nucl.\
Phys.\ {\bf B} 496, 231 (1997); S.~S.~Gubser, I.~R.~Klebanov, and A.~A.~Tseytlin, Nucl.\ Phys.\ {\bf B} 499, 217 (1997).


\bibitem{Brustein:2008cg}
  R.~Brustein and A.~J.~M.~Medved,
  Phys.\ Rev.\  D {\bf 79}, 021901 (2009); N.~Banerjee and S.~Dutta,
  JHEP {\bf 0903}, 116 (2009); M.~F.~Paulos,
  arXiv:0910.4602 [hep-th].

\bibitem{Iqbal:2008by}
  N.~Iqbal and H.~Liu,
  Phys.\ Rev.\  D {\bf 79}, 025023 (2009).


\bibitem{theoremcrosssection}
S.~R.~Das, G.~H.~Gibbons, and S.~D.~Mathur, Phys.\ Rev.\ Lett.\ {\bf 78}, 417 (1997); R.~Emparan, Nucl.\ Phys.\ B {\bf 516}, 297 (1998).

\bibitem{causalityreview} G.~S.~Denicol, T.~Kodama, T.~Koide and Ph.~Mota,
  J.\ Phys.\ G {\bf 35}, 115102 (2008); P.~Romatschke,
  arXiv:0902.3663 [hep-ph].

\bibitem{Koide:2009sy}
  T.~Koide, E.~Nakano and T.~Kodama,
  Phys.\ Rev.\ Lett.\  {\bf 103}, 052301 (2009).

\bibitem{taupiads} M.~P.~Heller and R.~A.~Janik,
  Phys.\ Rev.\  D {\bf 76}, 025027 (2007); R.~Baier, P.~Romatschke, D.~T.~Son, A.~O.~Starinets and M.~A.~Stephanov,
  JHEP {\bf 0804}, 100 (2008); M.~Natsuume and T.~Okamura,
  Phys.\ Rev.\  D {\bf 77}, 066014 (2008)
  [Erratum-ibid.\  D {\bf 78}, 089902 (2008)].







\end{thebibliography}
\end{document}